\documentclass[usenatbib]{mn2e}  
\usepackage{natbib}
\usepackage{epsfig}

\catcode`\@=11
\def\gta{\ifmmode{\,\mathrel{\mathpalette\@versim>\,}}
    \else{$\,\mathrel{\mathpalette\@versim>}\,$}\fi}
\def\lta{\ifmmode{\,\mathrel{\mathpalette\@versim<\,}}
    \else{$\,\mathrel{\mathpalette\@versim<}\,$}\fi}
\def\@versim#1#2{\lower 2.9truept \vbox{\baselineskip 0pt \lineskip
    0.5truept \ialign{$\m@th#1\hfil##\hfil$\crcr#2\crcr\sim\crcr}}}
\catcode`\@=12  
\renewcommand{\[}{\begin{equation}}
\renewcommand{\]}{\end{equation}}
{\newif\ifnotend
\notendtrue
\def\veclist{ABCDEFGHIJKLMNOPQRSTUVWXYZabcdefghijklmnopqrstuvwxyz.}
\def\top#1#2.{#1}
\def\tail#1#2.{#2.}
\loop\expandafter\xdef\csname v\expandafter\top\veclist\endcsname%
{{\noexpand\bf\expandafter\top\veclist}}
\edef\veclist{\expandafter\tail\veclist}
\if\veclist.\notendfalse\fi\ifnotend\repeat}
{\newif\ifnotend
\notendtrue
\def\veclist{ABCDEFGHIJKLMNOPQRSTUVWXYZ.}
\def\top#1#2.{#1}
\def\tail#1#2.{#2.}
\loop\expandafter\xdef\csname c\expandafter\top\veclist\endcsname%
{{\noexpand\cal\expandafter\top\veclist}}
\edef\veclist{\expandafter\tail\veclist}
\if\veclist.\notendfalse\fi\ifnotend\repeat}

\def\df{{\sc df}}

\def\fracj#1#2{{\textstyle{#1\over#2}}}
\def\pa{\partial}

\def\kms{\,{\rm km}\,{\rm s}^{-1}}

\def\kpc{\,{\rm kpc}}

\def\d{{\rm d}}

\def\figref#1{Fig.~\ref{#1}}
\newcommand{\beq}{\begin{equation}}
\newcommand{\eeq}{\end{equation}}

\title[Actions for axisymmetric potentials]
{Actions for axisymmetric potentials}

\author[James Binney]{James  Binney\thanks{E-mail:
binney@thphys.ox.ac.uk}\\
Rudolf Peierls Centre for Theoretical Physics, Keble Road, Oxford OX1 3NP, UK\\
}

\begin{document}

\date{Draft, June 28, 2012}

\pagerange{\pageref{firstpage}--\pageref{lastpage}} \pubyear{2012}

\maketitle

\label{firstpage}

\begin{abstract}
We give an algorithm for the economical calculation of angles and actions for
stars in axisymmetric potentials. We test the algorithm by integrating orbits
in a realistic model of the Galactic potential, and find that, even for
orbits characteristic of thick-disc stars, the errors in the actions are
typically smaller than 2 percent. We describe a scheme for obtaining actions by
interpolation on tabulated values that significantly accelerates the process
of calculating observables quantities, such as density and velocity moments,
from a distribution function.
\end{abstract}

\section{Introduction}

When electronic computers first became widely available, it was discovered
that orbits in typical axisymmetric galactic potentials usually admit three
isolating integrals of motion \citep{HenonH,Ollongren}. Consequently, by
Jeans' theorem, the distribution functions (\df s) of equilibrium
axisymmetric galaxies should be functions of three integrals of motion.
Unfortunately, analytic forms of all three integrals are known only for
exceptional potentials, so the few three-dimensional galaxy models in the
literature that have a known \df\ \citep[e.g.][]{Rowley} employ only the
classical energy and angular-momentum integrals $E$ and $L_z$, and therefore
lack generality.

The action integrals $J_r$, $J_z$ and $L_z$ are particularly useful constants
of motion \citep[e.g.][]{Canary}, and we have previously argued the merits of
models in which the distribution function is an analytic function of $J_r$,
$J_z$ and $L_z$. To take advantage of these models one should be able to
evaluate economically the actions of a star from its conventional phase-space
coordinates $(\vx,\vv)$. To date we have used two techniques for evaluating
actions: (i) torus construction \citep{KaasalainenB,BinneyM} and (ii) the
adiabatic approximation \citep{B10,BinneyM,SchoenrichB12}. Torus construction
is a general and rigorous technique and for some applications it is the
technique of choice \citep[e.g.][]{McMillanB12}. For other applications it is
inconvenient because it delivers $(\vx,\vv)$ as functions of the actions and
angles, rather than the actions and angles as functions of $(\vx,\vv)$.

The adiabatic approximation delivers actions and angles as functions of
$(\vx,\vv)$ but it is reasonably accurate only for stars that stay close to
the Galaxy's mid-plane. Here we introduce a different approximate way to
obtain actions, which, though still approximate, is more accurate than the
adiabatic approximation and is valid for stars that move far from the mid
plane.

\section{The algorithm}\label{sec:alg}

Our algorithm is based on the idea that the Galaxy's gravitational potential
is similar to a St\"ackel potential -- for a detailed description of the
latter see \cite{deZeeuw83}. St\"ackel potentials for oblate bodies are
framed in terms of prolate confocal coordinates. The latter are defined by
the distance $2\Delta$ between the foci of the coordinate curves. These foci
lie at $R=0$ and $z=\pm\Delta$, where $(R,z,\phi)$ is a system of cylindrical
polar coordinates. Following Binney \& Tremaine (2008; hereafter BT08)
\S3.5.3 we define new coordinates $(u,v)$ by
 \[
R=\Delta\sinh u\sin v\quad;\quad
z=\Delta\cosh u\cos v.
\]
 The generating function of the canonical transformation between these
systems of coordinates is
 \[
S(p_R,p_z,u,v)=p_RR(u,v)+p_zz(u,v)
\]
 so from $p_u=\pa S/\pa u$ we have
\begin{eqnarray}\label{eq:puv}
p_u&=&\Delta(p_R\cosh u\sin v+p_z\sinh u\cos v)\nonumber\\
p_v&=&\Delta(p_R\sinh u\cos v-p_z\cosh u\sin v).
\end{eqnarray}
 In these coordinates a St\"ackel potential can be written in terms of two
functions of one variable, $U(u)$ and $V(v)$, being given by
 \[
\Phi_{\rm S}(u,v)={U(u)-V(v)\over\sinh^2u+\sin^2v}.
\]
 This being so, the Hamilton--Jacobi equation yields (BT08 eq.~3.249)
\begin{eqnarray}
{p_u^2\over2\Delta^2}&=&E\sinh^2u-I_3-U(u)-{L_z^2\over2\Delta^2\sinh^2u}\nonumber\\
{p_v^2\over2\Delta^2}&=&E\sin^2v+I_3+V(v)-{L_z^2\over2\Delta^2\sin^2v},
\end{eqnarray}
 where $E$ is the orbit's energy and $I_3$ is a constant of separation.
 These equations make $p_u(u)$ and $p_v(v)$ functions of only their conjugate
coordinate, so we can evaluate the actions as
 \[\label{eq:Jrz}
J_r={1\over\pi}\int_{u_{\rm min}}^{u_{\rm max}}\d u\,p_u(u)\quad;\quad
J_z={2\over\pi}\int_{v_{\rm min}}^{\pi/2}\d v\,p_v(v),
\]
 where $u_{\rm min}\le u_{\rm max}$ are the roots of $p_u(u)=0$ and $v_{\rm
min}$ is the root of $p_v(v)=0$. Note that an orbit's actions are independent
of any system of coordinates and the subscripts $r$ and $z$ on the actions
merely remind us that, in a general way, $J_r$ quantifies oscillations
inwards and outwards, while $J_z$ quantifies oscillations around the
equatorial plane.

In as much as  our potential $\Phi$ is similar to a St\"ackel potential, we have
 \[
(\sinh^2u+\sin^2v)\Phi(u,v)\simeq U(u)-V(v).
\]
 Consequently, we have
\begin{eqnarray}\label{eq:deltas}
\delta U&\equiv&(\sinh^2u+\sin^2v)\Phi(u,v)-(\sinh^2u_0+\sin^2v)\Phi(u_0,v)\nonumber\\
&\simeq&U(u)-U(u_0)\nonumber\\
\delta V&\equiv&\cosh^2u \Phi(u,\pi/2)-(\sinh^2u+\sin^2v)\Phi(u,v)\\
&\simeq& V(v)-V(\pi/2).\nonumber
\end{eqnarray}
 Here $u_0$ is a reference value of $u$, the choice of which will be
discussed below, and the right side of the first equation appears to be a
function of $v$ but its dependence on $v$ will be weak unless $\Phi$ is very
unlike a St\"ackel potential. Similarly,
we assume that the dependence of the right side of the second equation on $u$
is at most weak. Then, given a point $(\vx,\vv)$ on the orbit we can
calculate two constants of motion:
 \begin{eqnarray}
I_3+U(u_0)&\simeq&I_3+U(u)-\delta U(u)\nonumber\\
&=&E\sinh^2u-{p_u^2\over2\Delta^2}-{L_z^2\over2\Delta^2\sinh^2u}-\delta
U(u)\nonumber\\
I_3+V(\pi/2)&\simeq&I_3+V(v)-\delta V(v)\\
&=&{p_v^2\over2\Delta^2}-E\sin^2v+{L_z^2\over2\Delta^2\sin^2v}-\delta
V(v).\nonumber
\end{eqnarray}
 Now we can evaluate $p_u$ for any given $u$ from
\[\label{eq:approx_pu}
{p_u^2\over2\Delta^2}\simeq E\sinh^2u-[I_3+U(u_0)+\delta
U(u)]-{L_z^2\over2\Delta^2\sinh^2u},
\]
 so we can evaluate the integral for $J_r$. The integral for $J_z$ is
evaluated in the same way.

In principle $u_0$ can be taken to be any quantity that is constant along an
orbit, but the accuracy of our work will depend on our choosing a value such
that the term in the definition (\ref{eq:deltas}) of $\delta U$ that contains
$u_0$ almost completely eliminates the $v$ dependence of the first term in
this equation. In fact, the natural choice for $u_0$ is the location
$\overline{u}$ of the minimum with respect to $u$ of $\delta U$ at fixed $v$.
This minimum can be determined before we have specified $u_0$ because the
derivative with respect to $u$ of the first of equations (\ref{eq:deltas}) is
manifestly independent of $u_0$. Physically $\overline{u}$ is the radial
coordinate of the shell orbit $J_r=0$ of given values of $E$ and $L_z$.

\subsection{Angle variables}

Equations (\ref{eq:puv}) for the momenta are obtained by solving the
Hamilton--Jacobi equation for the generating function
$S(u,v,\phi,J_r,J_z,L_z)$ of the canonical transformation  between the
$(u,v,\phi,p_u,\ldots)$ and the $(\theta_r,\theta_z,\theta_\phi,J_r,\ldots)$
systems of canonical coordinates with $S$ of the form
 \[
S=S_u(u,J_r,J_z,L_z)+S_v(v,J_r,J_z,L_z)+\phi L_z.
\]
 Given that $S$ takes this form, we may write
\begin{eqnarray}
S&=&\int\d u\,{\pa S_u\over\pa u}+\int\d v\,{\pa S_v\over\pa v}+\int\d\phi\,{\pa
S\over\pa \phi}\nonumber\\
&=&\int\d u\,p_u+\int\d v\,p_v+\phi L_z
\end{eqnarray}
 Hence 
\begin{eqnarray}
\theta_r&=&{\pa S\over\pa J_r}=\int\d u\,{\pa p_u\over\pa J_r}+\int\d v\,{\pa
p_v\over\pa J_r}\nonumber\\
\theta_z&=&{\pa S\over\pa J_z}=\int\d u\,{\pa p_u\over\pa J_z}+\int\d v\,{\pa
p_v\over\pa J_z}\\
\theta_\phi&=&{\pa S\over\pa L_z}=\int\d u\,{\pa p_u\over\pa L_z}+\int\d v\,{\pa
p_v\over\pa L_z}+\phi.\nonumber
\end{eqnarray}
 We obtain the  derivatives of $p_u$ and $p_v$ from the chain rule. For
 example
 \begin{eqnarray}\label{eq:puJr}
{\pa p_u\over\pa J_r}&=&{\pa p_u\over\pa E}{\pa E\over\pa J_r}+{\pa p_u\over\pa
I_3}{\pa I_3\over\pa J_r}\nonumber\\
&=&{\pa p_u\over\pa E}\Omega_r+{\pa p_u\over\pa
I_3}{\pa I_3\over\pa J_r},
\end{eqnarray}
 where $\Omega_r=\pa E/\pa J_r$ is the radial frequency, so
 \begin{eqnarray}
{\theta_r\over\surd2\Delta}&=&\Omega_r
\left(\int_{u_{\rm min}}^u\d u\,{\sinh^2u\over
p_u}+\int_{v_{\rm min}}^v\d v\,{\sin^2 v\over p_v}\right)\nonumber\\
&&-{\pa I_3\over\pa J_r}\left(\int_{u_{\rm min}}^u{\d u\over
p_u}-\int_{v_{\rm min}}^v{\d v\over p_v}\right).
\end{eqnarray}
 A detail possibly worth noting is that we always take $p_u$ of $p_v$ to be
given by the positive square root and when considering a point in phase space
at which $p_u<0$ we obtain the indefinite integrals over $u$ as twice the
corresponding integral from $u_{\rm min}$ to $u_{\rm max}$ minus the integral
from $u_{\rm min}$ to $u$ with $p_u$ taken to be positive. When this
procedure is followed for all integrals, the angle variables increase along
an orbit continuously as they should.

 The derivatives with respect to $J_r$ in equation (\ref{eq:puJr}) can be
obtained by observing that by the chain rule the matrix
\[
\pmatrix{\Omega_r&\Omega_z&\Omega_\phi\cr
\pa I_3/\pa J_r&\pa I_3/\pa J_z&\pa I_3/\pa L_z\cr
0&0&1}
\] 
 is the inverse of the matrix\footnote{Care must be taken with derivatives
with respect to $L_z$ regarding whether they are at constant $(E,I_3)$ or
$(J_r,J_z)$.}
 \[
\pmatrix{\pa J_r/\pa E&\pa J_r/\pa I_3&\pa J_r/\pa L_z\cr
\pa J_z/\pa E&\pa J_z/\pa I_3&\pa J_z/\pa L_z\cr
0&0&1}.
\]
 The latter is readily obtained by differentiating equations (\ref{eq:Jrz}) and
leads to  the definite integrals mentioned in the previous paragraph. 

\subsection{Interpolation}

To recover the observable properties of a model stellar system at a given
spatial point, such as its density $\rho$ and velocity dispersion tensor
$\sigma^2_{ij}$, one has to integrate the distribution function over all
velocities. These integrals entail large numbers of evaluations of the \df,
and it is important to keep down the cost of each evaluation. This goal
motivates us to tabulate the values of $J_r$ and $J_z$ as functions of the
classical integrals $E$, $L_z$ and $I_3+U(u_0)$ or $I_3+V(\pi/2)$. However,
$I_3+U(u_0)$ proves ill-suited to
this task because its numerical value varies rapidly as one moves through
action space. A more convenient constant of motion is
 \begin{eqnarray}\label{eq:Eu}
E_r&\equiv&{p_u^2\over2\Delta^2}
+{L_z^2\over2\Delta^2}\left({1\over\sinh^2u}-{1\over\sinh^2u_0}\right)+\delta
U(u)\nonumber\\
&&\qquad-E(\sinh^2u-\sinh^2u_0).
\end{eqnarray}
 At $u=u_0$, which we have chosen to be the minimum of the potential that
governs the motion in $u$, $E_r=p_u^2/2\Delta^2$ so we can think of $E_r$ as
the energy invested in radial oscillations. Consequently, for any values of
$E$ and $L_z$, $E_r$ vanishes for $J_r=0$ and takes its largest value for
$J_z=0$ and we can readily obtain $J_r$ and $J_z$ by interpolating between
the values taken by $J_r$ and $J_z$ at a grid of values of $E_r$. 

In detail we structure the grid in $(L_z,E,E_r)$ space as follows. The grid
points in $L_z$ are defined by the angular momenta of circular orbits with
radii uniformly distributed between minimum and maximum radii. For each value
of $L_z$ we adopt as grid points in  $E$ the energies
 \[
E_i=E_c(L_z)+\left({i\over2N}v_{\rm max}\right)^2,
\]
 where $E_c(L_z)$ is the energy of the circular orbit with angular momentum
$L_z$ and $\fracj12v_{\rm max}^2$ is slightly smaller than the difference
between the energy of that orbit and the escape energy from its circle.  For
each such energy we identify $u_0=\overline{u}$, the minimum with respect to
$u$ of 
\[
E\sinh^2u-\delta U-L_z^2/(2\Delta^2\sinh^2u).
\]
 Then we find the speed $v$ that the star
has at this spatial point and determine the values taken by $E_r$,
$I_3+V(\pi/2)$, $J_r$ and $J_z$ at the phase-space point
$(\vx,\vv)=(\Delta\sinh(u_0),0,v\cos\psi,v\sin\psi)$ for values of
$\psi$ uniformly distributed in $(0,\pi/2)$. With this scheme interpolation
errors can be kept below $\sim1\%$ with a grid of size $60\times50\times50$,
which takes $\sim30\,$sec to compute on a laptop.

The present algorithm lends itself to tabulation better than the adiabatic
approximation because with the present algorithm it is straightforward to
resort to the algorithm whenever actions are required for values of the
integrals that lie outside the grid. By contrast, when the adiabatic
approximation is used, values of $E_z$ are required for given $J_z$ and these
are hard to obtain beyond the limits of the pre-computed table of values of
$J_z$ for given $E_z$.

\begin{figure}
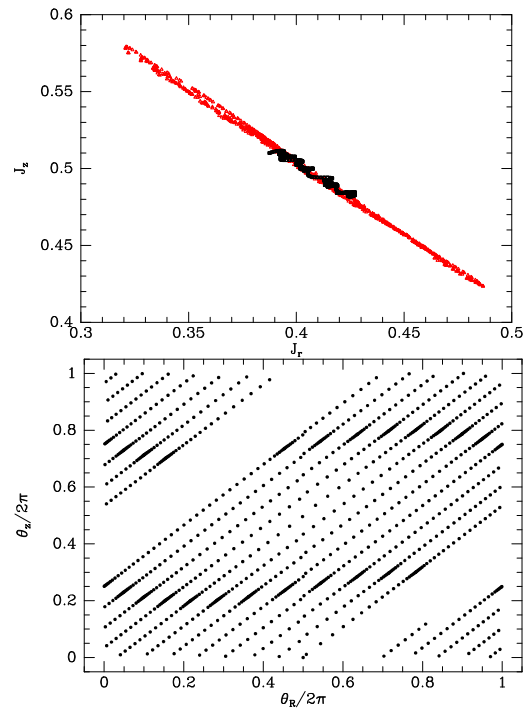

\centerline{\epsfig{file=c5/jujv.ps,width=.8\hsize}}
\centerline{\epsfig{file=c5/theRz.ps,width=.8\hsize}}
\caption{Top: values of $J_r$ and $J_z$ recovered along an orbit in a realistic
Galactic potential. The black points are obtained with the algorithm of
Section \ref{sec:alg} using $\Delta=3.5\kpc$ while the red points are
obtained with the adiabatic approximation. The units are $100\kms\kpc$. Bottom: the evolution of the angle
variables along this orbit.}\label{fig:JJ}
\end{figure}

\section{tests}

We have tested the algorithm by numerically integrating orbits in a realistic
Galaxy potential and after each time-step using the above algorithm to
determine $(\theta_r,\theta_z,J_r,J_z)$. Any variation in the recovered
values of the actions along the orbit quantifies errors in the procedure, as
do deviations of the motion in the $(\theta_r,\theta_z)$ lane from straight
lines. The adopted potential is that of model 2 of \cite{DehnenB} modified to
give the thin disc a scale height of $0.3\kpc$ -- this potential is generated
by exponential thin and thick stellar discs, plus a gas disc, an axisymmetric
bulge with axis ratio $0.6$ and a dark halo with axis ratio $0.8$. The upper
panel of \figref{fig:JJ} shows values of the actions along an orbit that has
corners at $(R,z)=(9.5,2)\kpc$ and $(6.6,1.35)\kpc$.  The black points are
obtained using the above algorithm, while the red points are obtained with
the adiabatic approximation in the superior formulation of
\cite{SchoenrichB12}.  Quantitatively, with the adiabatic approximation the
standard deviations of $J_r$ and $J_z$ are $(4.13,3.89)\kms\kpc$ while with the
above algorithm they are $(1.16,0.97)\kms\kpc$, smaller by a factor $\sim4$. The
lower panel shows the values taken by $(\theta_r,\theta_z)$ at each
integration step. The points lie on straight lines as required and the slopes
of plots of $\theta_i$ versus time agree accurately with the frequencies that
are recovered from the formulae of Section \ref{sec:alg}.

\begin{figure}
\centerline{\epsfig{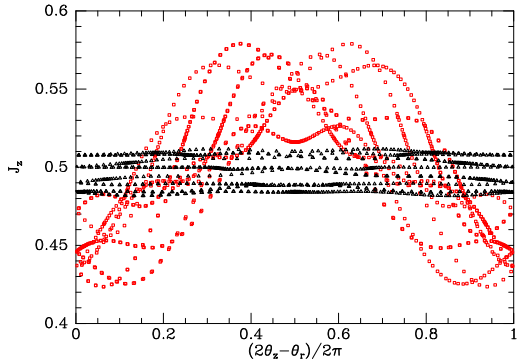}}
\caption{$J_z$ in units of $100\kms\kpc$ versus the combination of angle variables
$2\theta_z-\theta_r$ along the orbit that gives rise to \figref{fig:JJ}. The
black triangles are obtained with the algorithm of Section \ref{sec:alg}
while the red squares are obtained with the adiabatic
approximation.}\label{fig:Jtheta}
\end{figure}

Since the upper panel of \figref{fig:JJ} shows that the actions we recover,
either by the present algorithm or from the adiabatic approximation, are
tightly correlated, it is natural to ask what else they are correlated with.
Their correlations with $R$ and $z$ prove to be extremely small (especially
in the case of the present algorithm), but the red squares in
\figref{fig:Jtheta} show that in the case of the adiabatic approximation
$J_z$ (and therefore $J_r$ also) is correlated with the combination of angle
variables $2\theta_r-\theta_z$.  This angular dependence implies that as one
moves over an orbital torus at constant radius, the error in $J_z$ has one sign in
the plane and another far from it, and that the magnitude of this pattern of
errors oscillates between pericentre and apocentre, changing sign somewhere
in between. The black triangles in \figref{fig:Jtheta} show that the present
algorithm yields more accurate actions largely by eliminating this angular
dependence.

\begin{figure}
\centerline{\epsfig{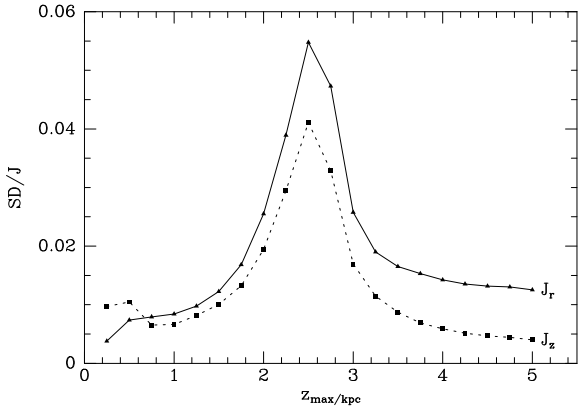}}
\caption{The ratios of the standard deviations in $J_r$ and $J_z$ to 
$(J_r+J_z)/2$ as functions of the maximum distance from the plane
attained on the orbit. Along this sequence of orbits $J_z$ rises from zero to
$240\kms\kpc$, while $J_r$ decreases from $50$ to $25\kms\kpc$.}
\label{fig:sd_juv}
\end{figure}

\figref{fig:sd_juv} plots the ratios of the standard deviations of $J_r$ and
$J_z$ to $(J_r+J_z)/2$ as functions of the maximum height $z_{\rm max}$ attained
on the orbit -- all orbits were started by dropping particles from
$(R,z)=(9.5\kpc,z_{\rm max})$. The fractional error in $J_z$ is never more
than 4\% and is rarely in excess of 2\%. The error in $J_r$ is larger but is
still generally under 2\% of the average action. The pronounced peaks in
the errors in both actions around $z_{\rm max}=2.5\kpc$ is probably connected
with the $1:1$ resonance between the horizontal and vertical motions: none of
the orbits contributing to the figure appears to be actually trapped, but for
$z_{\rm max}\sim2.6\kpc$ the frequency $\Omega_r-\Omega_z$ is very low.
Consequently, the small difference between $\Phi$ and a St\"ackel potential
has appreciable time to disturb the orbit.

\begin{figure}
\centerline{\epsfig{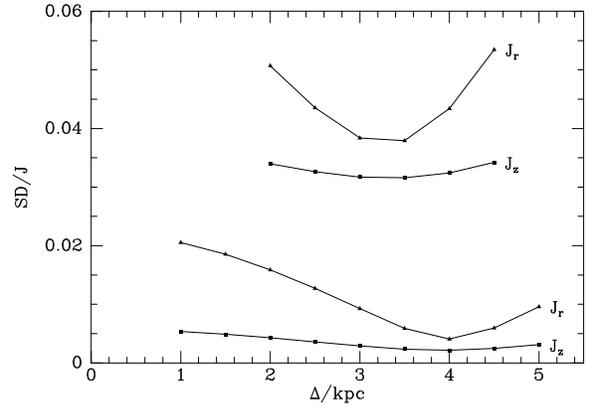}}
 \caption{The points above $\hbox{SD/J}=0.03$ show standard deviations of
$J_r$ (triangles) and $J_z$ (squares), normalised by $(J_r+J_z)/2$, along the
orbit that yielded \figref{fig:JJ} as functions of the value of $\Delta$ used
in the algorithm.  The lower points show the corresponding numbers for an
orbit that has its outer corner at $(R,z)=(9.5,1)\kpc$ rather than
$(9.5,2)\kpc$. The latter orbit has actions
$(45.7,15.6)\kms\kpc$.}\label{fig:sigvsD}
\end{figure}

The results shown in Figs.~\ref{fig:JJ} to \ref{fig:sd_juv} were obtained
with $\Delta=3.5\kpc$.  \figref{fig:sigvsD} shows the standard deviations of
$J_r$ and $J_z$ along two orbits as functions of $\Delta$. The orbits have
similar eccentricities, but different values of $z_{\rm max}$: the upper
squares and triangles are associated with an orbit that has $z_{\rm max}=2\kpc$,
while the lower triangles and points are for an orbit that has
$z_{\rm max}=1\kpc$. Both orbits have corners at $R\sim9.5$ and $\sim6.5\kpc$. We see
that the standard deviation in the values of $J_z$ along the orbit is much
less sensitive to the value of $\Delta$ than is the standard deviation of the
$J_r$ values. 

\section{Conclusions}

We have shown that values of actions and angles accurate to a couple of
percent can be obtained for orbits in a realistic axisymmetric model of the
Galactic potential by treating the potential as if it were a St\"ackel
potential. For orbits typical of observed stars belonging to either the thin
or thick discs the error in $J_z$ is always less than $\sim4\%$ of the average
action and is usually significantly smaller. The errors in $J_r$ are always
less than 6\% and usually less than 2\% of the average action.  Even in the
era of Gaia it is unlikely that the errors in the measured phase-space
coordinates of any star will be small enough that the inaccuracies inherent
in our algorithm will dominate the final uncertainties in derived angles and
actions.  The errors in actions obtained from the adiabatic approximation are
larger by a factor $\sim4$ for thin-disc stars and significantly larger still
for thick-disc stars.

A possibility that we have not pursued, but which might be important if one
needs to model an entire galaxy rather than the extended solar neighbourhood,
is to make the inter-focal semi-distance $\Delta$ a function of $L_z$ and $E$
-- by integrating a few orbits at wide-ranging values of $L_z$ and $E$ it
should be possible to choose a suitable functional form for $\Delta(L_z,E)$.

Each action evaluation requires a one-dimensional integral and with the
existing code takes $\sim100\,\mu{\rm s}$ on a laptop. Each angle evaluation
takes about twice as long because it requires of order two one-dimensional
integrals. Since evaluation of the observables that follow from a \df\
requires a great many evaluations of the actions, it is  cost-effective
to tabulate $(J_r,J_z)$ as functions of the classical integrals $(L_z,E,I_3)$
and we have described an effective scheme for doing this. In a companion
paper we illustrate what can be achieved using this scheme by fitting \df s
to observational data for our Galaxy.

\label{lastpage}
\end{document}